\documentclass[12pt]{article}
\usepackage{epsf}
\def\1ad{\mbox{\normalsize $^1$}}                                  
\def\2ad{\mbox{\normalsize $^2$}}                                  
\def\3ad{\mbox{\normalsize $^3$}}                              
\def\4ad{\mbox{\normalsize $^4$}}                                  
\def\5ad{\mbox{\normalsize $^5$}}                                  
\def\6ad{\mbox{\normalsize $^6$}}                                  
\def\7ad{\mbox{\normalsize $^7$}}                                  
\def\8ad{\mbox{\normalsize $^8$}}                                  

\def\npb#1#2#3{{ Nucl. Phys.} {\bf B#1} (#2) #3 }
\def\plb#1#2#3{{ Phys. Lett.} {\bf B#1} (#2) #3 }
\def\prd#1#2#3{{ Phys. Rev. } {\bf D#1} (#2) #3 }
\def\prl#1#2#3{{ Phys. Rev. Lett.} {\bf #1} (#2) #3 }

\def\ijmpa#1#2#3{{ Int. J. Mod. Phys.} {\bf A#1} (#2) #3 }

\def\jhep#1#2#3{{ J. High Energy Phys.} {\bf #1} (#2) #3 }

\def\bb#1{{\tt hep-th/#1}}

\def\CA{{\cal A}}   
\def\CL{{\cal L}}

\def\dj{\hbox{d\kern-0.347em \vrule width 0.3em height 1.252ex depth
-1.21ex \kern 0.051em}}

\def\half{{1\over 2}\,}

\def\Tr{{\rm Tr\,}}

\def\ket{\rangle}
\def\bra{\langle}

\def\btheta{\overline \theta}

\def\pt{\partial}

\newcommand{\nn}{\nonumber}

\newcommand{\ie}{\mbox{{\em i.e.~}}}

\def\Tr{\mbox{Tr}}
\def\bra{\langle}
\def\ket{\rangle}

\def\shalf{{\mbox{$\half$}}}                                         
\def\Dirac{\,\raise.15ex\hbox{/}\mkern-13.5mu D}                            
\def\dirac{\,\raise.15ex\hbox{/}\kern-.57em \partial}                  
\def\pslash{\,\raise.15ex\hbox{/}\kern-.57em p}                        
\def\Aslash{\,\raise.15ex\hbox{/}\kern-.57em A}                        
\def\epslash{\,\raise.15ex\hbox{/}\kern-.57em \varepsilon}         
\def\kslash{\,\raise.15ex\hbox{/}\kern-.57em k}
\def\nslash{\,\raise.15ex\hbox{/}\kern-.57em n}

\def\Fhat{{\widehat F}}                                                 
\def\Bhat{{\widehat B}}                                              
\def\Ahat{{\widehat A}}       
\def\bepsilon{\mbox{\boldmath $\varepsilon$}}                            
\def\bgamma{\mbox{\boldmath $\gamma$}}                             
\def\btheta{\mbox{\boldmath $\theta$}}                             
 
\newcommand{\be}{\begin{equation}}
\newcommand{\ee}{\end{equation}}
\newcommand{\ben}{\begin{equation*}}
\newcommand{\een}{\end{equation*}}
\newcommand{\ba}{\begin{eqnarray}}
\newcommand{\ea}{\end{eqnarray}}
\newcommand{\ban}{\begin{eqnarray*}}
\newcommand{\ean}{\end{eqnarray*}}
\newcommand{\brr}{\begin{array}}
\newcommand{\err}{\end{array}}
\newcommand{\bc}{\begin{center}}
\newcommand{\ec}{\end{center}}

\begin{document}

\newcommand{\sheptitle}                                                         
{Non-linear Vacuum Phenomena in Non-commutative QED}             
\newcommand{\shepauthora}                                                       
{{\sc                                                                          
 L. Alvarez-Gaum\'e and J.L.F.~Barb\'on}                        
\footnote[1]{On leave from
Departamento de F\'{\i}sica
de Part\'{\i}culas. Universidad de Santiago de Compostela, Spain.}
}

\newcommand{\shepaddressa}
{\sl
Theory Division, CERN \\
 CH-1211 Geneva 23, Switzerland \\
{\tt luis.alvarez-gaume@cern.ch}\\
{\tt barbon@mail.cern.ch}}


                           
\newcommand{\shepabstract}                                        
{We show that the classic results of Schwinger on the exact
propagation of particles in the background of constant field-strengths
and plane waves can be readily extended to the case of nocommutative QED.
It is  shown that non-perturbative effects on constant backgrounds are the same
as their commutative counterparts, provided the on-shell gauge invariant
dynamics is referred to a non-perturbatively related space-time frame. 

For the case of the plane wave background, we find evidence of the effective
extended nature      of non-commutative particles, producing  retarded and
advanced  effects
in scattering. Besides the known `dipolar' character of non-commutative
 neutral particles, we find that  charged particles
 are also effectively extended, but they   behave  instead as  
`half-dipoles'.
}

\begin{titlepage}
\begin{flushright}                                                       
CERN-TH/2000-181\\                                                                                                                
{\tt hep-th/y0006209}\\    
                                                      
\end{flushright}                                     
\vspace{0.5in}                                           
\begin{center}                                                    
{\large{\bf \sheptitle}}                                            
\bigskip\bigskip \\ \shepauthora \\ \mbox{} \\ {\it \shepaddressa} \\
\vspace{0.5in}                                                                 
                                                      
{\bf Abstract} \bigskip \end{center} \setcounter{page}{0}
 \shepabstract              
\vspace{1.5in}   
\begin{flushleft}                                                        
CERN-TH/2000-181\\                                                       
June 2000                  
\end{flushleft}                                       
                                                     
                                                                    
\end{titlepage}

\newpage                                                           

                                                                  
\subsection*{Introduction and Conclusions}

  In a
classic paper \cite{schw}, J. Schwinger solved, to leading order in $\hbar$ and to
all orders in the coupling constant $e$, the QED non-linear effects
on particle propagation in  a couple of  particular background
configurations:
constant
field strengths, and plane waves. Using the proper time method, the classical
non-linearities can be exactly summed at tree level, defining
  exact Green's functions
 for charged particles in these backgrounds. One can
obtain  from
these `dressed propagators' various  important physical quantities such
as dispersion relations, the Euler--Heisenberg effective
 Lagrangian to all orders,
and the decay of a coherent electric field by pair production.

   This paper is devoted to  a generalization of  Schwinger's
results to the same background configurations in
 non-commutative QED. Although we mostly consider tree-level effects,
our results are non-perturbative in the coupling constant, thus providing some
non-perturbative information on the dynamics of the non-commutative theory.

At a purely perturbative level, non-commutative field theories (NCFTs)
 are defined by simply replacing standard
products of fields by their  Moyal products \cite{pertbunch}:
\be
\psi(x)\star\varphi(x) =
 {\rm exp}\left({i\over 2} \theta^{\alpha\beta} {\partial
\over \pt\xi^\alpha} {\pt\over \pt\eta^\beta}\right)
 \,\psi(x+\xi) \,\varphi(x+\eta)
\Big |_{\xi=\eta=0}
.\ee
This introduces a peculiar non-local structure on
 length scales of $O(\sqrt{\theta})$
due to the non-commutativity of the coordinates \cite{conn} 
\be
\label{comm}
[x^{\alpha}, x^\beta ] =i\,\theta^{\alpha\beta}.
\ee

From the physical point of view, the most important effect of the non-local
Moyal product is to give particles an effective extension, depending on
the amount of energy-momentum
 they carry in the conjugate directions, in the sense of
 eq. (\ref{comm}).
 This leads to a topological classification of  Feynman diagrams
in NCFTs,  \cite{planar}
and  to many other properties  that are reminiscent of
open-string dynamics. Indeed, one of the main interests of these theories
is as toy models of string dynamics, without the complications of the
gravitational sector. The precise relation between NCFTs and string theory
involves backgrounds with large Neveu--Schwarz $B$-field fluxes, a much-studied
subject recently \cite{cds, bunch, SW, ampli, elec}. 

One of the unexpected surprises of the perturbative studies was the lack of
decoupling of non-commutative effects in generic theories \cite{uvir, uvirs},
 at least
within perturbation theory. In particular, the precise way in which
 the NCFTs negotiate
their infrared singularities at a non-perturbative level is an important 
open question. Another interesting surprise was the discovery of inconsistencies
in NCFTs when time is involved in the non-commutativity \cite{elec, br},
unless the effects of (\ref{comm}) are masked by stringy fuzziness.
For instance, a concrete 
 problem is the lack of unitarity of the $S$-matrix at a perturbative level
\cite{jaume}.
There is also evidence that the same problems persist at the level of the
large-$N$ master field \cite{br}. 

In this context, it is thus  interesting to study some non-perturbative
effects in a characteristic  NCFT, such as the non-commutative version of
QED. One of the distinctive non-perturbative effects described by Schwinger,
the decay of a constant electric field via pair production, has a string
analog with special features, namely the decay rate diverges at a critical
value of the electric field \cite{bachaspor}.
 This is the same critical field that governs
the classical singularity of the Born--Infeld effective action
 \cite{burgess}. 
 The relation
between electric fields in open-string theory and time/space non-commutativity
makes this example even more interesting. One of our chief objectives will be
the search for similar features in  NCFT.

The propagation of charge-$q$ particles                                   
 in the background of a gauge field $A(x)$ is controlled              
by the                                                                
 spectral properties of the covariant-derivative  operators           
 $D^{(q)}_{A \star}$ with                                                     
\ba                                                                       
\label{covdev}
D^{(1)}_{A(x) \star} &=& \partial_x +ieA(x)\star , \nn \\     
D^{(0)}_{A(x) \star} &=& \partial_x +ie[\star A(x)\star, \;]            
\ea                                                                       
for the cases of neutral $q=0$ matter, photons and ghosts, and the    
usual charged $q=1$ matter, such as Dirac fermions or charged scalars.         
 In particular, we are interested in the effective world-line Hamiltonian
\be                                                                
 H(D^{(q)}_{A\star})= -(D^{(q)}_{A\star} )^2 + (V^{(q)}_{\rm eff})\star,
 \ee
with $V^{(q)}_{\rm eff}$ a spin- and charge-dependent
 effective potential, linear in   the non-commutative field strength $\Fhat$:
\be
\label{ncfs}
\Fhat_{\mu\nu} =  \partial_\mu A_\nu - \partial_\nu A_\mu +ie \,A_\mu
\star \, A_\nu  - ie\,A_\nu \star\,A_\mu.
\ee
  For scalar and vector fields, $H$ is the
 kinetic kernel of  fluctuations in the $A(x)$ background. For spin $\half$
particles,                                                            
  the effective world-line Hamiltonian $H(\Dirac_{A\star})$   is          
given by                                                                
\be                                                                             
H(\Dirac_{A\star}) = (i\Dirac_{A\star})^2 - m^2=                                       
  -(D_{A\star})^2 -m^2 - {e\over 2}\,\sigma_{\mu\nu}{\widehat               
F}^{\mu\nu}\star                                                               
,\ee                                                                            
and is related to the Dirac operator by the non-commutative generalization      
of the  usual relation                                                   
\be                                                                
{1\over i\Dirac_{A\star} -m} =
 {i\Dirac_{A\star} +m \over (i\Dirac_{A\star})^2 -m^2} =
(i\Dirac_{A\star} +m ) \,{1\over H(\Dirac_{A\star})}
.\ee                                                                 

We wish to study the Green's function for $H(D_{A\star})$, 
\be                                                                             
\label{gf}
G_{A\star}\,                             
(x,y)=\bra x |{1\over H(D_{A\star})} |y\ket                                    
,\ee 
or  `dressed' 
propagator to all orders in the coupling constant (a different notion
of non-commutative Green's function was developed in \cite{nikig}).                                                                            
We shall  also discuss 
 the one-loop                                                      
`effective action' (in practice, various subtractions are             
needed in its definition):                                              
\be                                                                       
\label{eac}
\Gamma [A\star] = 
 (-1)^{2J+1}\,\half\, {\rm Tr}\,{\rm log}\,H(D_{A\star})              
,\ee   
for each field of spin $J$ in the matter and gauge sectors, including the ghosts. In
this equation the trace `Tr' is
 over the space-time, spin and internal quantum
numbers.                                                                       
                                                                               
Both the dressed Green's function and the effective action can be        
obtained   from the heat kernel $K_{A\star}(x,y;s)$,  defined by  
the equations                                                             
\be                                           
\label{hker}
\left({d\over ds} + H(D_{A\star})\right) \,K_{A\star} (x,y;s) = 0        
,\qquad                                                                     
\lim_{s\to 0} K_{A\star}(x,y;s) = \delta(x-y)                            
,\ee                                                                            
using the standard formal expressions:   
\ba                                                                        
\label{hkern}
G_{A\star}\,(x,y) &=& \int_0^\infty ds \;\bra x|                                
e^{-sH(D_{A\star})}|y\ket=                                                     
\int_0^\infty ds\,K_{A\star} (x,y;s),                                 
\nn \\                                                                
&\,&\nn\\                                                                 
\Tr\,{\rm log}\,H(D_{A\star})
 &=&\int dx \int_0^\infty {ds\over s} \,{\rm tr} \,\bra
 x|e^{-sH(D_{A\star})}  
|x\ket                                                                  
 =  \int_0^\infty {ds\over s} \int dx \,{\rm tr}\,                             
K_{A\star} (x,x;s),                                                            
\ea                                                                            
where `tr' is now a trace over internal and spin quantum numbers.

 At a
technical level, the non-commutative problems (8-11) can be mapped
to commutative problems with appropriate effective background
configurations that are still exactly solvable.  From the
physical point of view, we find rather similar physical effects,
depending on the non-commutative deformation parameter $\theta$ only
through the non-commutative field strength $\Fhat_{\mu\nu}$, provided
the space-time quantum numbers are properly interpreted.

 It is important to stress that we are not
making use of the mapping to commutative variables of ref. \cite{SW}. Our
mapping to commutative variables is tailored to the problem of determining
the dynamics of a probe particle in our restricted class of backgrounds.

In the case of constant field-strength background, the identification
of good gauge invariant quantum numbers is non-trivial. This is due to the
known fact that non-commutative gauge
 theories have no strictly local gauge-invariant
operators. Although covariant local
 expressions are easy to devise, the construction
of the gauge-invariant trace involves an
 integration over spacetime, neglecting
surface terms at infinity, a problematic procedure for constant fields,
since they do not turn-off at infinity.  

We find that asymptotic on-shell quantities, such as dispersion relations,
 can be defined unambiguously,
as well as integrated quantities over all spacetime, such as
the real and imaginary parts of the effective {\it action}.
 This is again reminiscent
of string theory, where the introduction of local sources is notoriously
difficult. The proper definition of these gauge invariant quantities involves
the use of a certain non-perturbative geometrical frame,
 different from the perturbative one. This is one of our
main results; the appropriate notion of `energy', as well as the causality
structure, depends on the background field strength and the non-commutative
 deformation parameter in a precise way. 

In the case of the plane-wave background we find that physical properties
are characterized by the extended nature
 of probe particles; the known `dipole'
structure of non-commutative quantum-field quanta \cite{dipoles}.
 We also find a novel phenomenon
for charged particles; namely there is a similar extensivity effect, with the
particles  behaving
as `half-dipoles'. The one-loop effective action in the plane-wave 
background is found to be trivial up to boundary terms. In the non-commutative
case, this result depends on a judicious choice of ultraviolet and 
infrared cutoffs, reflecting the UV/IR connection pointed out
in \cite{uvir, uvirs}.

\subsection*{Constant Field-Strength Background}

Consider the family of field configurations of the form 
\be
\label{linans}
A_\mu (x)   
 = -\half                                                            
F_{\mu\alpha} x^\alpha =-\half (Fx)_\mu,
\ee
 with $F_{\mu\nu}$ {\it not necessarily} antisymmetric.   
The corresponding non-commutative 
field strength (\ref{ncfs}) is given by 
\be                                     
  {\widehat F}_{\mu\nu} = 
\label{fhateq}
\left(F_A
-{\frac{e}{4}} F\theta F^t\right)_{\mu\nu},
\ee
where $F_A$ denotes
 the antisymmetric part of $F$ and $F^t$ is the transpose.
For example,
if we consider a fully aligned magnetic $2\times 2$ block, such that
 $F$ has a magnetic flux
on the same plane as $\theta^{\mu\nu}$ or, in other words
 ${\bf B} \parallel {\btheta}_m$,
we                         have, restricting to that plane: 
\be                                                                        
\label{genan}                                                                   
F=                                                                         
 \left(\begin{array}{cc} S_1 & B
 \\ -B & S_2 \end{array}
\right),                                                               
\qquad \theta =                                                                 
 \left(\begin{array}{cc} 0 & \theta_m \\ -\theta_m & 0 \end{array}
\right),                 
 \qquad {\widehat F} = \left(\begin{array}{cc} 0 & {\widehat B} \\ -{\widehat B
}   
 & 0\end{array}
\right)                                             
,\ee  then the non-commutative magnetic field ${\widehat
B}$ is related to the parameters in $A_\mu$ and $\theta$ by                                
\be                                                                  
 {\widehat B} =
B-{e\over 4} \,\theta_m \left(B^2 +S_1 S_2\right).    
\ee                                                                  
  We see that, for
$\theta\neq 0$, the
symmetric part of
$F$ does contribute
non-trivially to
$\Fhat$, unlike the commutative case. In fact,
including the
symmetric part of
$F$ in the gauge
potential is
essential in 
obtaining non-commutative magnetic fields of arbitrary magnitude and sign.

An important property of (\ref{fhateq}) is its quadratic nature, so 
 that different gauge potentials can lead  
         to the same ${\widehat
F}$, \ie  
 we can `complete the square' and write 
\be                             
\label{cuadrado}         e\,{\widehat  F} +{1\over \theta} =  {1\over
\theta}                    \,\left(1+{e\over
2}\theta\,                                                     F\right) \left(1-{e\over 2}
\theta\,F^t \right)                    = \left(1+{e\over 2}F\,\theta\right)
\left(1-{e\over 2}F^t \,\theta\right) \,{1\over
\theta}.                                                                    \ee 
 For
instance, if  $F=F_A$ is antisymmetric, 
 we have two solutions (in fact more) corresponding to $F$ and $F'$, related by:
\be    \label{dual}                                
1+{e\over 2} F\,\theta = -1-{e\over 2} F'\,\theta.           
\ee                                                                  
 In particular, a non-zero antisymmetric $F$ satisfying                      
$                                                                     
1+{e\over 2} F\,\theta =0                                                       
$                                                                      
corresponds to a 
 vanishing non-commutative field strength $\Fhat=0$. Therefore, it is
natural  
to
expect that these configurations are pure gauge. In this respect, we  point
out           
                    that the antisymmetric part of
$F$ is not the commutative field strength $F_{\rm
SW}$                                        
 of                                                                             
the Seiberg--Witten mapping \cite{SW}. The precise relation is              
\be                                                                             
\label{swrel}
e\,{\widehat  F} +{1\over \theta} =  {1\over \theta}                           
\,\left(1+{e\over 2}\theta\,
F\right) \left(1-{e\over 2} \theta\,F^t \right) =
{1\over \theta \left({1\over \theta} -e\,F_{\rm SW}                             
\right)\,\theta}    .\ee                                        
Notice that the mapping
 $\Fhat (F_{\rm SW})$ is one-to-one if we interpret the 
previous                                                              
equation for arbitrary field strengths in the sense of analytic continuation.
Thus, from the point of view of the Seiberg--Witten mapping, $\Fhat=0$ is
equivalent to $F_{\rm SW}=0$, when written in commutative variables, and
should characterize the vacuum unambiguously.  
    
\subsubsection*{Gauge Ambiguity of the Linear Ansatz}                                                                               
The existence of discrete symmetries of eq. (\ref{ncfs}) that look like
gauge transformations motivates a systematic discussion of the 
 gauge ambiguity of the linear ansatz
(\ref{linans}). Appart from the trivial gauge freedom of adding a constant
to the gauge potential,             
    let us find out the conditions for two gauge potentials of
the form     
\be                                                                             
\label{gpot}                                                                    
A=-\half\,F\,x, \qquad A^g = -\half\,F^g \,x,                                   
\ee                              to  actually be
 gauge-equivalent, namely  the existence of  a
star-unitary             transformation $g(x)$
satisfying                                       
\be                                                                             
\label{gauge}                                                  
    A^g = g^{-1}  \star
A\star g -{i\over e} \, g^{-1}\star\partial\,g,              
\ee                                                                             or,
star-multiplying by $g(x)$ on the left: 
$$                                              
ie\,g\star \,A^g = \left(\partial +ie\,A\star\right)\,g                         
.$$                                        
Notice that, at this point, 
  we are not assuming  any particular form for  $F$ and $F^g$; 
  they are not antisymmetric in general.                                  
                                                                     
Plugging in the linear ansatz  (\ref{gpot}) and using the formulas           
for left-right star products:                              
\ba                                                                             
\label{products}
x^\mu \star g(x) &=& \left(x^\mu + {i\over 2} \theta^{\mu\nu}\partial_\nu\right)
\,g(x), \nn \\                                                                  
g(x) \star x^\mu &=& \left(x^\mu - {i\over 2} \theta^{\mu\nu}\partial_\nu\right)
\,g(x),                                                                         
\ea                             
 we arrive at                                                                   
\be                                                                    
\label{lineq}                                                                   
\left(1+{e\over 4} (F+F^g)\,\theta\right)\,\partial g = -i{e\over 2} \,
(F-F^g)\,x\,g.                                                                  
\ee                                                                            
 Assuming that the matrix in the left hand side is invertible, we can
write this equation in the form                  
\be                                                                             
\left(\partial +ie\,A_{\rm eff}\right)g=0,     
\ee                                                                        
with                                                                            
\be                                                                  
\label{aeff}                                                                 
A_{\rm eff} (x)
 = -\shalf \,F_{\rm eff}\,x \equiv-\shalf \, {1\over 1+{e\over 4}(   
F+F^g)\theta}\,(F^g -F) \,x.                                                    
\ee                                                                             
Namely, the gauge transformation is covariantly constant (in the ordinary       
sense) with respect to a linear potential given by (\ref{aeff}). The            
function $g(x)$ is solved as a Wilson line in $A_{\rm eff}$ but,  in order      
to really define a gauge-transformation function, it should be independent
of the integration path. This only happens if the associated field-strength     
{\it vanishes}. In other words,  we need                               
\be                                                                             
\left({1\over 1+{e\over 4}(                                            
F+F^g)\theta}\,(F^g -F)\right)_{\rm antisym} =0.                                
\ee                                                                            
Assuming, again, that the matrix in the denominator is non-singular, and noting
the antisymmetry  $\theta$, this condition is equivalent to
$$                                                                         
F_A -{e\over 4}\,F\,\theta\,F^t = (F^g)_A -{e\over 4}\,F^g\,\theta\,(F^g)^t     
.$$                                                                            
Thus, we obtain the expected result that two linear potentials are         
gauge-equivalent when they have the same field strengths:                       
$                                                                             
{\widehat F} (F) = {\widehat F}(F^g)                                            
$.                                                                       
The associated  gauge transformation may be written  as  
\be                                                                    
\label{gtrans}
g(x) = {\rm exp}\left({ie\over 4} \,x \,(F_{\rm eff}) \,x + \dots \right) 
,\ee                                                                    
where the dots represent a trivial constant and a possible linear term, that
simply adds a constant to the gauge potentials. 
Restricting now to   antisymmetric $F$, we seem to have found  that             
$F$ and $F'$ related by (\ref{dual}) are gauge equivalent.                     
 However, the ${\bf Z}_2$ transformation $F\rightarrow F'$ is                 
{\it singular} as a gauge transformation since, precisely for
$F^g = F'$,  one has   
$$                                                              
1+{e\over 4} (F+F')\,\theta =0,                                            
$$                                                                            
and the effective connection $A_{\rm eff}$ in (\ref{aeff}) is not defined.
Thus, if we restrict ourselves to {\it non-singular} gauge transformations,
we would be forced to conclude
 that the ${\bf Z}_2$ transformation is not            
gauge and both branches are physically different.
 According to this interpretation 
of our formulae, there would be a physical  `screening effect' consisting       
on the neutralization of $\Fhat$ by the $\theta$-background, a conclusion
that would be at odds with our previous considerations based on
 the Seiberg--Witten mapping.      On the
other
hand, it is easy to see that the ${\bf Z}_2$ transformation (\ref{dual}) is
a member of a continuous set of regular gauge transformations that do not
change $\Fhat$. To be more specific, consider the magnetic aligned ansatz
of eq. (\ref{genan}), with the restriction $S_1=S_2=S$ for
simplicity.      
Then, the potentials that are gauge-equivalent to the vacuum $\Bhat=0$ lie on a circle 
of radius
$2/e\theta_m$ in                                                                        the
$(B,S)$ plane, centered at the point $(B=2/e\theta_m, S=0)$.  There
is                                        
    a gauge transformation of the form
(\ref{gtrans}) that takes
the vacuum at  $S=B=0$  into any of the points lying
 on the $\Bhat =0$ circle.                                                     
Of these, the gauge transformation that takes $B=S=0$ to $ S^g=0,             
 B^g=4/e\theta_m$                                              
 is singular. It is also the only singular
one, since the determinant                                                   
$$
{\rm det}\left(1+{e\over 4}F^g\,\theta\right) = \left(B^g
- {4\over e\theta_m} \right)^2 + \left(S^g \right)^2 = 0              
$$
only vanishes for $S^g =0, B^g = 4/e\theta_m$.
 Namely, the singular gauge transformation
is a                                   
                         member of a continuous
{\it regular} family of gauge transformations. 
Since                                                  
              it can be
approximated by honest gauge transformations and it does not introduce
 singularities                                                                  
in physical, gauge-invariant quantities, it should be allowed. It should
be interesting to study this question from the point of view of the
Wilson-loop operators defined in \cite{japon}.

\subsubsection*{Reduction to the Commutative Frame}               
   
The basic property of the linear ansatz (\ref{linans}), allowing us to solve
the problem for  constant fields, 
is the general form                    of the Moyal product by a single
variable:                            
\be           
 \label{prods}
  x^\mu \star
= x^\mu + {i\over 2} \theta^{\mu\nu}\partial_\nu  \qquad  
           \left[x^\mu
\star,\;\right] = i\theta^{\mu\nu}\partial_\nu                      
.\ee    
Using these relations we can map the basic covariant derivatives
 (\ref{covdev})
into ordinary differential operators. It is convenient to distinguish the
cases of charged, $q=1$, and neutral, $q=0$, probe particles.

\subsubsection*{\underline {\sl Charge One}}

Direct application of the first equation in (\ref{prods}) leads to
$$                                                                             
\partial_x +ieA(x)\star = \partial_x                                            
 -{ie\over 2} F\,x\star = \left(1+{e\over 4} F\theta                            
\right) \partial_x -{ie\over 2}F\,x                                             
.$$                                                                            
It is convenient to define the new  coordinates         $x'^\mu$
by the relation                         
\be                                                                              x^\nu =
x'^\mu  \left(1+{e\over 4} F\,\theta\right)_\mu^{\;\;\nu}               
.\ee                                                                            Using the
antisymmetry of  $\theta$, we can write the transformation          
 matrix                                                        
in appropriate form for left-action
 on contravariant vectors, $x=M \,x'$ with        
\be                                                               
\left(M_{(q=1)}\right)^\mu_{\;\;\nu}                                       
 =\left(1-{e\over 4}\theta\, F^t\right)^\mu_{\;\;\nu}                           
.\ee                                                                 
Finally, we apply the identity  
\be                                                                             
{\widehat F} = F_A -{e\over 4} F\,\theta \,F^t = F\,M_1 - F_S                   
,\ee                                                                            
with $F_S$ the symmetric part of $F$,                                           
obtaining  the following
 reduction formula to a purely commutative operator:       
\be                                                                             
\label{madremuno}                                                               
D^{(1)}_{A(x)\star} \equiv 
\partial_x +ieA(x)\star =  \partial_{M_1^{-1} x} +ie {\widehat
A}(M_1^{-1} x)   
\equiv
D_{\Ahat(x')}
,\ee    
with   the equivalent commutative gauge potential given by:    
\be                                                                             
\label{equi}                                                                     {\widehat
A}(x') = -\half ({\widehat F} + F_S) \,x'                              
.\ee                                   
                                        Thus, up to
the expected replacement $F\rightarrow {\widehat F}$,
 we have just a linear transformation
of the coordinates. Notice that, once we have a commutative operator,
the                                               
symmetric part  $F_S$ in the previous formula is a pure $U(1)$
 gauge ambiguity.  We  stress that
the                                                   reduction formula is only valid
provided $M^{-1}$ exists.                    

The full non-commutative gauge symmetry (\ref{gauge}),
 acting through Moyal products:
\be
D_{A\star} \rightarrow D_{A^g \star} = g^{-1} \star D_{A\star} \star g,
\ee
is mapped after the reduction to an ordinary $U(1)$ action on $\Ahat$, plus
a coordinate transformation on the $x'^\mu$. Indeed, the transformation matrix
$M_1$ is not gauge invariant, as one can easily check for example using the
aligned magnetic ansatz of (\ref{genan}) (we set $e=4$ to simplify the
 notation)
\be                                                                             
\label{muno}                                                                 
M_1 = 1- \theta\,F^t =                                                     
 \left(\begin{array}{cc} 1-\theta_m\,B  & -\theta_m\,S_2     
 \\ \theta_m\,S_1 & 1-\theta_m\,B \end{array}    
\right).                                                               
\ee                                                                             
 Not even the determinant         
$$                                                                             
{\rm det}(M_1) = (1-\theta_m\,B)^2 
+\theta_m^2 \,S_1 S_2 = \half -\theta_m\,\Bhat -
\theta_m\,     
\left(B-{1\over 2\theta_m}\right)                                               
$$                                                                             
is gauge invariant. Notice 
that $M_1$ is singular at the locus of ${\rm det}(M_1)=0$,
or
\be
{\rm det}\left(1-{e\over 4}\,\theta\,F^t \right) =
{\rm det}\left(1+{e\over 4}\,F\,\theta \right)=0.
\ee
Hence, the singularities of the transformation to the commutative frame coincide
(for the $q=1$ case) with the singular gauge transforms of the vacuum.

\subsubsection*{\underline{\sl Charge
Zero}} 
                                                                                 The
neutral covariant derivative admits a similar reduction          
            with the
commutative covariant derivative being trivial                         
\be                                                                             
\label{madremcero}                                                              
D^{(0)}_{A(x)\star}\equiv \partial_x +ie[\star A(x)\star,\;] = \partial_{M_0^{-1}
x}                            
,\ee                                                                            
 and the transformation                                                         
matrix differing by a rescaling of the coupling:                                
\be                                                                             
\label{mcero}                                                                   
M_{(q=0)} = 1 -{e\over 2} \theta\,F^t                                           
.\ee                                                                           
   In this
case the
transformation
matrix $M_0$
is almost
gauge
invariant.                                                                     
Namely, one can rewrite (\ref{cuadrado}) in the form                            
$$                                                                            
1+e\,\theta\,\Fhat = \theta\,(M_0)^t\,\theta^{-1}\,M_0 = 
 (M_0)^t \,\theta^{-1} \, 
M_0\,\theta = 1+e\,\Fhat\,\theta                                               
,$$    
so that,                                                                        
\be                                                                           
\left({\rm det}(M_0)\right)^2 = {\rm det}\left(1+e\,\theta\,\Fhat\right)        
\ee                                                                             
is manifestly  gauge-invariant (the individual entries of $M_0$ are, 
however, not gauge-invariant). Thus, in this case
  the singular locus of the transformation $x\rightarrow x'$, satisfying 
\be
{\rm det}\left(1-{e\over 2}\,\theta\,F^t \right) =0={\rm det}\left(1+e\,\theta\,
\Fhat
\right), 
\ee
is gauge-invariant. It is interesting that the singularity coincides
 with that of the Seiberg--Witten 
mapping to commutative field strengths $F_{\rm SW}$, given in eq.
 (\ref{swrel}).

\subsubsection*{The Green's Functions}                                   
                                                                                
We can use these results to write a reduction formula for
 the dressed propagator.
 From the defining equation
   \be     
H(D_{A\star}) \,G_{A\star}(x,y) = \delta(x-y)                            
,\ee                                                                          
we get                (at this
point, it is not necessary to
specify the charge $q$)   
\be                                                                             
\label{greenf}                                                                  
G_{A\star} (x, y) = {1\over |{\rm det}(M)|} G_{\widehat A} \,( M^{-1} x, M^{-1} 
y)                                                                              
,\ee                                                                              where the
determinant factor comes from the transformation of the delta  
        function, and
$G_{\widehat A}$ is the Green's function of the
 commutative problem    for the       
field ${\widehat A}$.            
        Taking
Fourier transforms:                                                 
$$                                                                             
{\widetilde G}_{A\star} (p) =  \int dx\, e^{ipx}\, G_{A\star}(x)               
 = \int {dx\over |{\rm det}(M)|}                                              
\,G_{\widehat A}\,( M^{-1} x) \,e^{ipx}                                  
= \int dx \,G_{\widehat A}\,( x)\, e^{ipMx}                                     
,$$                                                                  
we obtain the momentum-space reduction formula:                            
\be                                                                             
\label{momred}                                                       
{\widetilde G}_{A\star} (p) = {\widetilde G}_{\widehat A}                    
\,(pM)                                                                          
.\ee                                                                            
                                                                                
Alternatively, we can obtain the same results                                   
 by considering of  eigenvalue problem                               
\be                                                                             
H(D_{A(x)\star}) \,\psi_n (x) = \lambda_n \,\psi_n (x)                             
.\ee               
 Our 
reduction algorithm gives                                                  
\be                                                                             
H( D_{\Ahat(x')} ) \,\psi_n (Mx') = \lambda_n \,\psi_n (Mx')                 
,\ee                                                                              with
$D_{\Ahat(x')}$ the corresponding commutative operator in the potential
 $\Ahat$ defined                                                              
by eq. (\ref{equi}).  
 Thus, the original and reduced operators actually have the same spectrum
of eigenvalues. 
 Denoting by ${\widehat \psi}_n (x)$ the normalized eigenfunctions
of $H( D_{\Ahat})$,                                                
 we obtain                                                                
   \be     
\psi_n (x) = |{\rm det}(M)|^{-\half} \, {\widehat \psi}_n (M^{-1} x),
\ee                                                                          
where the precise proportionality constant comes from the unit normalization of 
the eigenfunctions. From here                                                   
we can write the  Green's function using its formal
  spectral definition:                         
\ba                                                                             
G_{A\star}(x,y) &=& \sum_n {\psi_n (x) \psi_n (y)^* \over \lambda_n }    
= \sum_n {1\over |{\rm det}(M)|} {                                              
{\widehat \psi}_n (M^{-1}x) {\widehat \psi}_n (M^{-1} y)^* \over \lambda_n }    
\nn \\                                                                          
 & =&                                                                           
{1\over |{\rm det}(M)|} \,G_{\Ahat}\,(M^{-1} x,  M^{-1} y)             
,\ea                                                                            
in agreement with  (\ref{greenf}).

A similar relation  follows for the                                   
heat kernel:                                                                    
\be                                                                  
\label{heatk}                                                                 
K_{A\star}\,(x,y;s)                                                             
 = {1\over |{\rm det}(M)|} \,K_{\widehat A}\,(M^{-1} x, M^{-1} y;s)             
.\ee

Explicit expressions for the Green's functions can
be written by applying the reduction formula  (\ref{greenf}) to
Schwinger's result in the case of charge-one
particles \cite{schw} (see also \cite{IZ}, pag. 100).
 The singularities of $G_{A\star} (x,y)$
at non-perturbative `large' values of the gauge
field $A(x)$ coincide with ${\rm det}(M_1)=0$ and
should be interpreted as gauge artifacts, since
we found that such gauge potentials are singular gauge transforms
of the vacuum.                                       
                                      
The Green's function for $q=0$  neutral particles                               
is the simplest possible; in momentum
space:                                                     
\be                                                                             
{\widetilde G}_{q=0} \,(p) = {i\over (pM_0)^2 - m^2 +i0}    
={i\over (p')^2 -m^2 +i0}.                           
\ee                                                                                         
It is interesting that the causal structure
inherited from the perturbative rule $m^2
\rightarrow m^2 -i0$ is the standard 
Feynman prescription, in terms of the rotated momenta $p'
= p M_0$. We shall provide some perspective
on this observation in the next section.

\subsubsection*{Dispersion Relations}

   The Green's functions found above are instrumental in constructing
 the perturbative 
expansion in the background of a  constant non-commutative field-strength
$\Fhat$. 
 However, non-linear 
 tree-level effects 
 on particle propagation  can  be  extracted directly from the gauge-invariant
 information 
contained in the singularities of the Green's function, \ie the 
dispersion relations,  
that  are                                                                 
nothing but the diagonalized version of the Klein--Gordon equations:            
\be                                                                             
H(D_{A\star})\,\varphi(x) = 0.    
\ee                                                                             
Thus, the analysis of dispersion relations can be carried out
  quite  generally  without 
explicit                                                                        
knowledge of the full form of the                                             
 Green's function. Using the general relation                      
\be                                                                             
{\widetilde G}_{A\star} (p) = {\widetilde G}_{\widehat A}                    
\,(pM)                                                                     
,\ee                                                                            
and the dispersion relation of the commutative problem:                   
\be                                                                          
f_{\Ahat} \,(p) =0,   
\ee                                                                             
as  determined by the singular locus of the 
  momentum-space Green's function
 ${\widetilde    
G}_A \,(p)$, or by                               
                         the solution of
Klein--Gordon-like equation:                                    
\be                                                                             
H(D_{\Ahat(x)})\,\phi(x) =0,   
\ee  
we can immediately
 write down the corresponding dispersion relation for the
 non-commutative   
counterpart. It is given by:                                                    
\be                                                                             
\label{nco}
f_{\widehat A}\,(pM)=0,                                                         
\ee                                                                           
where $M$ is either $M_0$ or $M_1$ depending on the  case we consider.
  In fact,
the correct form of the dispersion relation is                    
\be                                                                             
\label{corrr}                                                                    f_{\Ahat}
\,(p') =0,                                                         
\ee                                                                             
with                                                                            
\be                                                                             
\label{pprima}                                                                   p'_\mu =
p_\nu \,M^\nu_{\;\mu}                                                  
.\ee                                                                             The
reason why (\ref{corrr}) is the correct form is that the momenta $p_\mu$,   
 conjugate to                                                                   
the original non-commutative 
coordinates $x^\mu$, are not good quantum numbers   
for non-zero $\Fhat$ and $\theta$.
 The dispersion relation cannot contain explicitly 
the transformation matrix $M$, because it is not gauge invariant.

 We can illustrate these ideas with a couple of examples.  If we      
deal with ordinary   charge-one particles in a magnetic field pointing in the
$z$-direction,                              
                       we know the dispersion
relation is independent of the momenta in the $(x,y)$-plane
 $p_x, p_y$. There is an                                                      
infinite degeneracy of energy levels, 
labeled by {\it one} momentum variable, plus a
discrete
energy       
spectrum of Landau levels.
 Namely the states $|p_x, p_y \ket$ are substituted, on
diagonalizing                                                         
  the
Hamiltonian, by the states $|p_y, n\ket$, with the energy depending only on 
 the
oscillator                      
  quantum
number $n$. 
Our  choice of $p_y$ as the 
 continuous momentum label in the $(x,y)$-plane is a  gauge-dependent choice, as any other
linear combination of $p_x$ and $p_y$ would be valid,  but the
spectrum                                           
                      is gauge-invariant. One
obtains, for a charge-one particle,      
\be                                                                      
 E^2 = m^2
+ p_z^2 + e|B| (2n+1+\alpha)                                          
,\ee                                                                       
 with
$\alpha$ related to spin quantum numbers, taking $2J+1$ values.  
          
 Therefore, the
non-commutative counterpart  according to (\ref{nco}) is     
\be                                                                             
\label{moregen}                                                                 
((pM_1)_0)^2 = m^2 + ((pM_1)_z)^2 + e|{\widehat B}| (2n+1 + \alpha)             
.\ee  
In gauge-invariant form: 
\be\label{moregenin}
(E')^2 = m^2 + (p'_z)^2 + e|\Bhat| (2n+1+\alpha)                                
.\ee 
This example illustrates how the background fields determine what
are the good quantum numbers of on-shell 
asymptotic states. Even in the commutatative case, we learn
that only one linear combination of $p_x, p_y$ is a good quantum number
for the problem. In the non-commutative case, the gauge-invariant
dispersion relation is written in terms of $E'$ and $p_z'$ which are
the good energy and momentum variables.                   In addition,
we have the discrete label of Landau levels $n$, and an infinite
degeneracy labeled by one linear combination of $p'_x$ and $p'_y$.                                                          
                                                                                
An even simpler example is given by the dispersion relation for neutral
particles.
In this case
the
 pole is determined by                                                   
$$                                                                           
(pM_0)_\mu \,\eta^{\mu\nu}\,(pM_0)_\nu = m^2                                    
.$$                                                                             
We can summarize this by defining an effective metric $G_{\rm eff}$:           
\be                                                                             
\label{eef}                                                                     
(G_{\rm eff})^{-1} = M_0 \,\eta^{-1}\,M_0^t = \left(1-{e\over 2}\theta\,F^t     
\right) \,\eta^{-1} \,\left(1+{e\over 2} F\,\theta\right)                       
,\ee  
which would give $M_0$ the interpretation of a {\it vierbein}.                
 The dispersion relation  reads:                                            
\be                                                                             
\label{fakedr}                                                               
(G_{\rm eff})^{\mu\nu} p_\mu p_\nu = m^2                                        
\ee                                                                             
Considering for example a purely antisymmetric $F$, one can readily check that  
the effective metric degenerates in an analog of the Born--Infeld singularities
of                                                                              
string theory in background electric fields. One can also detect superluminal
propagation, in the sense of $dE/d|{\bf p}| >1$, for specific values
of the background fields.      In any case, such effective  
metric is not gauge invariant. Considering  the general magnetic-aligned        
 ansatz (\ref{genan})                                                           
with $e=4$ one
finds    
for the $2\times 2$ magnetic block of the effective inverse metric:   
$$
 (G_{\rm
eff})^{-1} = -M_0 (M_0)^t =                                           
 -\left(\begin{array}{cc} (1-2\theta_m\,B)^2 +4\theta_m^2\,S_2^2
  & 2\theta_m\,(
1-2\theta_m \,B)(S_1-S_2)                                                       
 \\  2\theta_m\,(                                                         
1-2\theta_m \,B)(S_1-S_2)  & (1-2\theta_m\,B)^2 +4\theta_m^2\,S_1^2 
 \end{array}
\right).                                                                        
$$                                                                            
Since the gauge-invariant combination of $B, S_1$ and $S_2$ is                 
$$                                                                             
|{\rm det}(M_0)| =
 1-4\theta_m\,\Bhat = (1-2\theta_m\,B)^2 + 4\theta_m^2\,S_1 S_2
,$$                                                                             
we see that the individual entries of the effective metric
 are not gauge invariant.
Therefore, even in the neutral case, we are led to defining
 new momentum quantum 
numbers for the problem in a deformed space:                                    
\be                                                                             
\label{rotas}
p' = p M                                                                        
.\ee    

Whenever the entries $M^{0i}$ of the transformation matrix are non-vanishing,
the frame-transformation advocated here mixes the energy and momentum
running in the free propagators of the 
 weak coupling expansion. This happens even in the case of pure
magnetic fields, provided the  $\theta^{0i}$ components linking the
time direction with the magnetic-flux plane are non-vanishing.  The
resummation of all tree interactions with the background, to all
orders in the electromagnetic coupling $e$, produces effective momentum
variables with standard causal structure.

\subsubsection*{Effective Action and Particle Production}                       
  
According to our previous considerations, the spectrum of the 
covariant-derivative operators does not change under the reduction operation, 
and  we expect
the one-loop determinant of these operators to be formally equal:
\be
{\rm det}\,H(D_{A(x)\star}) = 
\prod_n \lambda_n = {\rm det}\,H(D_{\Ahat(M^{-1} x)}).
\ee
Alternatively, using the general transformation rule for the heat kernel 
(\ref{heatk})
 we can read-off
 the effective action from the general expression (\ref{hkern}).  One obtains
\ba
\Gamma[A\star] &=& {(-1)^{2J+1}\over 2}
 \int_0^\infty {ds\over s} \int dx\, K_{A\star}\;
(x,x;s) \nn \\
 &=& {(-1)^{2J+1} \over 2} \int_0^\infty {ds\over s} \int dx\,|{\rm
det}(M)|^{-1}\; K_{\Ahat} \;(M^{-1} x,M^{-1} x;s)
.\ea
Changing variables to the $x'$-frame yields 
\be
\label{goodw}
\Gamma[A\star] =\half (-1)^{2J+1}
\int_0^\infty {ds\over s} \int dx'\, K_{\Ahat}\; (x',x';s)
= \Gamma[\Ahat],
\ee
which is precisely Schwinger's result for
 the field-strength $\Fhat$.  Notice that there is
a certain ambiguity in this manipulation. Had we decided to use translational
invariance of the heat kernel in (\ref{heatk})
 (\ie the fact that it 
 only depends on the
difference $x-y$) we would have obtained a different result
 for the effective
Lagrangian, by a factor of $|{\rm det}(M)|^{-1}$. 
 The presence of such a factor
would be rather problematic, since it is
 not gauge invariant for $q=1$ particles.
This factor would multiply the probability density of pair production (the 
imaginary part of $\Gamma$), as well as
the logarithmic divergence of the real part of $\Gamma$. In the first case
it would lead to a non-gauge-invariant decay rate. In 
the second case we would learn
that, upon expanding in a power series in the coupling $e$, 
 the theory needs an
infinite number  of high-derivative counterterms of the form $(\theta \Fhat)^n$ to
subtract logarithmic divergences. This would be at odds with the diagrammatic
analysis of ultraviolet divergences, where one finds that switching on $\theta$
always results in an improved ultraviolet behaviour, rather than the contrary.

The difficulties in defining a gauge-invariant {\it integrand}, or effective
{\it Lagrangian}, as opposed to just defining an effective {\it action},
 is another
manifestation of the lack of standard local gauge-invariant operators in these
theories. The ambiguity mentioned above is likely to be related to the 
difficulties
in handling total derivatives for constant fields.  In particular, 
if we consider
an adiabatic situation with a slowly varying field-strength,
 it is clear that
the heat kernel is not translationaly invariant any more. 

On the other hand, once we write the answer in the $x'$-frame as in  
(\ref{goodw}),
 since
we are actually integrating over $x'$, and they are dummy variables, 
 we may as
well replace them by the original coordinates $x$. Thus we find a
 gauge-invariant
density of decay probability given by Schwinger's result \cite{schw,IZ}:
\be
\label{finali}
w(x) = (-1)^{2J+1} \,(2J+1)\,{\alpha_{\rm em} |{\widehat E}|^2 \over 2\pi^2}\,
\sum_{n=1}^{\infty}
 {(-1)^{n(2J+1)} \over n^2} \,e^{-(n\pi m^2 / |e{\widehat E}|)}
,\ee
where $J$ is the spin of the particle pairs that are being produced.

      \subsection*{ Plane-Wave Background}                                                
     In this
section we
consider
the      
problem of
particle
propagation in
the background
of a plane
wave. For
simplicity, we
take it to be
linearly
polarized:  
\be
\label{pwave}
   A^\mu (x) 
= \varepsilon^\mu \,A(\xi)                                           
,\ee           
 with $\xi
= n^\mu x_\mu$, $n^2 = \varepsilon \cdot n =0$,                       
 $\varepsilon^2 =-1$.                                                           
It is easy to see that this background satisfies the classical equations
of motion of non-commutative QED. 

We use the notation $(x^\mu) = (t, x, y^1, y^2)\equiv (t,x, {\bf
y})$ and  choose coordinates so that $\theta$ is skew-diagonal with non-trivial
entries
\be\label{convu}
[x,y^1]=i\theta_m, \qquad [t,y^2]=i\theta_e.
\ee
Using the remaining $SO(1,1)\times SO(2)$ symmetries of the problem,
 we can let the wave propagate in the $(x,t)$-plane:  
\be
\label{conv}
n=(1,1,0,0), \qquad \varepsilon =(\varepsilon^0, \varepsilon^0,                  \bepsilon
), \qquad \bepsilon^2 =1                                              
.\ee   
An important simplification is that the field
strength equals the commutative
one:                                                          
\be                                                                      
\Fhat_{\mu\nu} = F_{\mu\nu}= f_{\mu\nu} \,A' (\xi) = (n_\mu \varepsilon_\nu -
n_\nu                                                                           
\varepsilon_\mu ) \, A' (\xi)                                                   
,\ee
where the prime stands for derivative with respect to $\xi=t-x$.
In the following we consider the particular case of $q=1$ Dirac fermions and
$q=0$ Majorana fermions, for the purposes of definiteness, although the results
are easily generalized to particles of other spin assignments. 

\subsubsection*{
The Equivalent
Commutative Problem}

The Dirac    operators $i\Dirac^{(q)}_{A\star}$,     
when acting on functions of the form                                            
$$                                                                            
\psi_{\bf k} (t,x) \,e^{i{\bf ky}}                                             
,$$                                                                             
become the standard-product operator                                            
\be\label{redop}       
i\dirac_{(t,x)} + \bgamma \cdot {\bf k} - m -e
 \epslash \CA^{(q)}_{\bf k}                    
,\ee     
with a shifted effective potential                                             
\be                                                                             
\label{caun}
\CA^{(q=1)}_{\bf k} =
 A(nx-\shalf n\theta k)                                            
\ee
for the charged  case, and                                                      
\be      
\label{cados}
\CA^{(q=0)}_{\bf k} =
 A(nx-\shalf n\theta k) - A(nx + \shalf n\theta k)                 
\ee     
for the neutral case, where $k=(0,0,{\bf k})$. The                            
 shifts act only on the null projection                                         
of $\theta {\bf k}$  on the $(t,x)$ plane. Therefore, it is appropriate
to use a mixed coordinate-momentum representation and analyze the
effective two-dimensional effective operator (\ref{redop}) at fixed
values of ${\bf k}$. 
                                     
 In the following, we just use    
$\CA_{\bf k} (\xi)$ as a unified notation,
 keeping in mind that it depends implicitly    
on $q$ and $\theta^{\mu\nu}$.   Notice that a carefully tuned periodic
wave can render the shift effects irrelevant for a particular value of
the transverse momentum ${\bf k}$.

\subsubsection*{Scattering and Advanced/Retarded Effects}                                           
The scattering of particles of charge $q=0,1$ by the plane wave can
be solved exactly by simply generalizing old textbook results (see
for instance \cite{IZ}). We wish to solve the non-commutative
Dirac wave equation
\be
\label{ncdwe}
(i\Dirac^{(q)}_{A\star}-m) \,\psi(x) =0,
\ee
with $A_\mu$ as in   (\ref{pwave}). We assume that $A(\xi)$ is a function
with compact support. This implies that in the asymptotic past $\psi(x)$
can be chosen to be an incoming monochromatic plane wave. With the choice
for $n^\mu$  (\ref{conv}) we know that ${\bf k}=(k^1,k^2)$ is conserved.
Scattering is purely one-dimensional and it will be reduced to computing
the phase shifts of the outgoing wave-function after the particle   traverses
the wave-front. We can look for solutions of (\ref{ncdwe}) of the form
\be
\label{trial}
\psi(x) = e^{-ik\cdot x} \,\psi_k (t,x)
,\ee
with $k^\mu = (k^t, k^x, {\bf k})$ and $k^2=m^2$ as implied by the asymptotic
conditions. After the reduction to the commutative operator (\ref{redop})
we have the ordinary equation:
\be
\label{ordin}
\left[i\dirac-e\,\epslash\,\CA_{\bf k} -m\right] \,e^{-ik\cdot x} \,
\psi_k (t,x) =0, 
\ee
where    $\CA_{\bf k}$ is given by   (\ref{caun}) for $q=1$
 and (\ref{cados}) for $q=0$. 
Note that, although $k$ is a four-momentum, in $n\theta k$ only ${\bf k}$
appears for $n=(1,1,0,0)$. 

When $q=1$, if $n\theta k >0$, the particle sees the pulse a time $\shalf 
|n\theta k|$ before the commutative counterpart would, whereas for 
$n\theta k <0$, it will interact with it later than the commutative counterpart
by the same amount $\shalf |n\theta k|$. 

The $q=0$ case is more curious. For commutative neutral
 particles there is absolutely
no interaction with the wave front. In the non-commutative case however the
particle sees a pulse $A(\xi-\shalf n\theta k)-A(\xi+\shalf n\theta k)$,
containing always and advanced and a retarded component of opposite signs.
Only when the transverse momentum ${\bf k}$ vanishes the interaction
disappears. 

This once again hints at the interpretation of particles in
non-commutative field theory as extended one-dimensional
 objects with charges at the endpoints. In the $q=0$ case we can
think of them as dipoles whose size depends on the transverse momentum
${\bf k}$, whereas for $q=1$ they seem to behave as `half-dipoles' also
with an effective size depending on ${\bf k}$ (perhaps in this case it should
be more appropriate to think of a dipole but with one end at infinity). 

The solution of (\ref{ordin}) is straightforward \cite{IZ} and we do not 
dwell on the details. Picking a solution of
the free Dirac equation $u(k)$ satisfying
$$
(\kslash -m) \,u(k) =0, \qquad {\bar u}\,u = 2m,
$$
 we can write the result as:
\be
\psi_k (t,x) = \left(1+{ie\nslash \epslash \over 2k\cdot n} \,\CA_{\bf k}
(\xi) \right) \,e^{iI_k} \,u(k),
\ee
where
\be
I_k = -k\cdot x - \int_{-\infty}^{n\cdot x}  d\xi \,\left(e{\CA_{\bf k}
(\xi) \,\varepsilon \cdot k \over n\cdot k} + e^2 \,{\CA^2_{\bf k} (\xi) \over
2n\cdot k} \right)
\ee
is the eikonal of the particle, if we  think in terms of geometrical
optics. The phase shift now depends on the value of  the transverse  momentum   
of the particle scattered, and it incorporates the advanced and/or retarded
effects of the wave front $\CA^{(q)}_{\bf k}$ due to the effective extended
nature of the particles.

\subsubsection*{The Green's Function}                                                                                 
For a more detailed description of the propagation in the background of
the plane wave we need to compute the Green's
function                                                            
\be
    S(x,x') = \bra x \vert {1\over i\Dirac_{A\star} -m} \vert x' \ket   
=                                                 
                       \int {d{\bf k}
\over (2\pi)^2} e^{i{\bf k}({\bf y}-{\bf y}')} \;S_{\bf k} (t,x   ;
t',x')                                                                        
,\ee                                                                            where the
mixed-representation Green's function is given by                     
\be                                                                            S_{\bf k}
(t,x; t',x') =                                                        
 \bra t,x | {1\over i\dirac_{(t,x)} + \bgamma \cdot {\bf k} - m -e              
\epslash      
\CA_{\bf k} } | t', x' \ket     
.\ee 
We shall compute it through the bosonic one:  
$$                                                                              
S_{\bf k} (t,x ; t',x') = \left(i\dirac_{(t,x)} +\bgamma \cdot {\bf k} - e      
\epslash \CA_{\bf k}
 + m \right) G_{\bf k} (t,x ; t',x')                                
,$$  
with                                                                            
$$                                                                              
G_{\bf k} (t,x ; t',x') = \bra t,x | {1\over                                    
 \left(i\dirac_{(t,x)} +\bgamma \cdot {\bf k} - e                               
\epslash \CA_{\bf k}  \right)^2 -m^2 } 
|t', x' \ket                                     
.$$ 
Following Schwinger's
treatment in the proper-time
method:                                                            
\be
\label{schwp}
    G_{\bf k}
(t,x ; t',x') = -i \int_{-\infty}^0 d\tau \,\bra t,x | e^{-i\tau       (H+i0)}
|t',x' \ket                                                      
\ee                                                                             where the effective world-line Hamiltonian is given by 
\be \label{effhal}
H= \pi_a \pi^a + V, 
\ee
 with the
definitions                                                          
\ba                                                                              \pi_a
&=&
i\partial_a - e \CA_a                                                   
 = i\partial_a -e \varepsilon_a \CA_{\bf k} (\xi),     
\nn\\       
V&=& -{\bf k}^2 - m^2 +2e {\bf k}\cdot       
 \bepsilon \,\CA_{\bf k} (\xi) -e^2 \CA_{\bf k} (\xi)^2
-                             
             {e\over 2} \sigma^{\mu\nu} f_{\mu\nu} \,
 \CA_{\bf k}'
 (\xi)                             
.\ea                                                                               In the
following, we use latin index notation for the two-dimensional         
   $(t,x)$-plane. Our goal is to evaluate the matrix element in (\ref{schwp})
explicitly solving the quantum mechanical system on the world-line. 
We have            the basic commutators                                               
\be                                                                              [\pi_a ,
x_b ] = i\eta_{ab} ,\qquad [\pi_a, \pi_b ] = -ie F_{ab} =0 ,\qquad      [\pi_a, V] = i
\partial_a V                                                     
.\ee                                                                               Notice
the vanishing of $F_{ab}$ in the $(t,x)$-plane. This represents      
     a rather
important simplification with respect to the direct four-dimensional world-line 
problem.                                                                                                                                                                                                                                      
The two-dimensional world-line  equations of motion
are:                                            
\ba                                                                              {dx^a
\over d\tau} &=& i[H,x^a ] = -2\pi^a, 
\nn
\\        
{d\pi^a \over d\tau} &=& i[H,\pi^a] =
 \partial^a V = n^a V' (\xi)                 
.\ea                                                                               Two
important identities that follow are   
\be                                                                              {d \over
d\tau} (\pi^a n_a) = 0, \qquad \left[ \xi , {d\xi \over d\tau}\right]   =
0                                                                             
,\ee
so that  the quantity 
\be                                                                              -2 \pi_a
n^a = {\xi(\tau)-\xi(0) \over \tau}                                    
\ee                                                                               is a
constant of the motion. Using this, we integrate the equation of        
    motion of
$\pi^a$:                                                            
\be    \label{pieq}                                                                           \pi^a =
D^a + {\tau \over \xi(\tau) - \xi (0)} \,n^a V                          
\ee                                                             
in terms of a constant operator $D^a$, commuting with $\pi^a n_a$. Using        
$\pi^a = -\shalf dx^a /d\tau$ and integrating further, we get                   
$$        
-\shalf \left(x^a (\tau) - x^a (0)\right) = D^a \tau + {1\over (2\pi^a n_a )^2} 
\,\int_{\xi(0)}^{\xi(\tau)} d\xi \;n^a V(\xi)                                   
.$$
From here we determine $D^a$, and plugging it back into the expression for      
$\pi^a$  (\ref{pieq}): 
\be       
\pi^a = -{x^a_\tau-x^a_0 \over 2\tau} +{\tau \over \xi_\tau -\xi_0} \,n^a V(\xi)
 - {\tau \over (\xi_\tau - \xi_0)^2}                                            
\,\int_{\xi_0}^{\xi_\tau} d\xi \,n^a V(\xi)                                     
.\ee  
With these elements we can write  $H$
 as a function of $x_\tau^a$ and $x_0^a$. In order to evaluate the
matrix element (\ref{schwp})  we need       
to order the $x_\tau^a$ chronologically (notice that $[\xi_\tau, \xi_0 ]=0$
and thus  we do not care about their ordering).  
The basic commutator is                                                         
$$                                                                           
[\xi_0 , x^a_\tau ]= [\xi_\tau + 2(\pi_b n^b)\,\tau, x^a ]= 2n_b \tau [\pi^b,   
x^a] = 2in^a\,\tau                                                              
.$$      
We also need the value 
of $[x^a (\tau), x_a (0) ]$, which can be obtained  by solving    
for $x_0^a$ as a function of $x_\tau^a$:                                        
$$         
x_0^a = x_\tau^a + 2\tau \,\pi_\tau^a + {\tau \over (\xi_\tau -\xi_0)^2} \,     
\int_{\xi_0}^{\xi_\tau} d\xi \,n^a V(\xi)                                       
.$$      
Hence
$$     
[x^a_0 , x_{a\tau} ] = 2\tau [\pi^a , x_a] + \left[                             
{\tau \over (\xi_\tau -\xi_0)^2}                                                
\int_{\xi_0}^{\xi_\tau}  d\xi \;V(\xi) , \xi \right] = 2\tau \cdot i\delta^a_a  
= 4i\tau                                                                    
.$$    
Now we can compute the world-line Hamiltonian directly from its definition      
$H=\pi^a \pi_a + V$ and order it into the form:                                 
\be      
H= {1\over 4\tau^2} \left(x^a (\tau)x_a (\tau) -2x^a (\tau) x_a (0) + x^a (0)   
x_a (0) \right) -{i\over \tau} +
 {1\over \xi_\tau -\xi_0} \int_{\xi_0}^{\xi_\tau
} d\xi \,V(\xi)                                                                 
.\ee      
The heat kernel takes the form      
\be        
U_{\bf k} (t,x; t',x', \tau) = \bra t,x | e^{-i\tau H}  |t',x' \ket =           
C_{\bf k} (t,x ; t',x') \;e^{-i\int^\tau d\tau' F_{\bf k} (t,x;t',x', \tau')}   
,\ee     
where                                                                           
\be       
F_{\bf k} (t,x;t',x',\tau) = {1\over 4\tau^2} \left[(t-t')^2 - (x-x')^2 \right] 
-{i\over \tau} + {1\over \xi-\xi'} \int_{\xi'}^{\xi} V                          
.\ee       
We find, upon integrating in $\tau$:  
\be                                                                             
U_{\bf k} (t,x;t',x',\tau) = {C_{\bf k} (x_a; x_a') \over \tau} \,              
e^{i\left[{(x_a -x_a')^2 \over 4\tau} -{\tau \over \xi-\xi'}\int_{\xi'}^\xi V   
\right]}                                                                        
.\ee      
The $\tau$-independent prefactor is fixed by requiring                          
\be                                                                             
(i\partial_a -e\CA_a ) U_{\bf k} (t,x;t',x', \tau) =                            
( -i\partial'_a -e\CA_a) U_{\bf k} (t,x;t',x', \tau)                            
=                                                                               
 \bra t_\tau, x_\tau |       
\pi_a |t'_0, x'_0 \ket                                                          
.\ee      
These equations imply that $C(x_a, x_a')$ is covariantly constant in each       
argument. Hence  it is given by a two-dimensioal 
Wilson line, with the overall scale fixed
by requiring                                                                    
$$                                                                            
\lim_{\tau\to 0} U_{\bf k} (t,x;t',x',\tau) = \delta(t-t')\,\delta(x-x')        
.$$    
Explicitly:   
\be    
C_{\bf k} (t,x;t',x') = -{1\over 4\pi} \,
 \Phi^{(2)}_{\bf k}                     
(t,x;t',x') = -{1\over 4\pi}\;                                                  
{\rm exp}\left(-ie\int_{(t',x')}^{(t,x) }                                       
dz^a \CA_a \right)                                                              
.\ee   
The Wilson line can be further reduced using $\varepsilon^a = \varepsilon^0 \,  
n^a$:                                                                           
\be \label{lw}    
\int_{(t',x')}^{(t,x)}                                                          
dz^a \CA_a = 
\varepsilon^0 \,\int_{\xi'}^{\xi} d{\bar \xi} \CA_{\bf k} ({\bar            
\xi})  
,\ee
where we
have
integrated
on a
straight
line.                                                                               
Evaluating the integral over the effective potential 
$ V$ and putting all factors together,
we get the following       expression for the four-dimensional bosonic
kernel:                             
\be     \label{bosk}                                                                         U(x_\mu ;
x'_\mu , \tau) = \int {d{\bf k} \over (2\pi)^2} U_{\bf k}(t,x;t',x',
\tau)      \;e^{i{\bf k}({\bf y}
- {\bf y}' )}                                             
,\ee 
where                                                                           
$$
  U_{\bf k}
(t,x;t',x',\tau) =
 -{1\over 4\pi\tau} \Phi^{(2)} (t,x;t',x') \;   
     e^{iF^{(2)}} \;{\rm 
exp}\left(i  
       \tau {\bf k}^2 -2i\tau
e{\bf k}\cdot \bepsilon                                  
 \int_{\xi'}^{\xi} {\CA_{\bf k} \over
\xi-\xi'}\right)
,$$
\be\label{ddf}
F^{(2)} (t,x;t',x',\tau)
 = {(x_a -x_a')^2 \over 4\tau} + \tau \,m^2 + \tau e^2  
\int_{\xi'}^{\xi} {\CA_{\bf k}^2 
\over \xi-\xi'} + \shalf \,e\tau \,f_{\mu\nu}\sigma^{  
\mu\nu} \;{\CA_{\bf k} (\xi)-\CA_{\bf k}(\xi') \over
\xi-\xi'}                                   
.\ee    
We can check this expression by comparing it with Schwinger's result            
for $\theta=0$, where $\CA_{\bf k}(\xi) = A(\xi)$.
 In this case              
the ${\bf k}$-integral is gaussian:                                             
$$                                                                             
\int 
 {d{\bf k} \over (2\pi)^2} e^{i{\bf k}({\bf y}-{\bf y}')} {\rm exp}\left(   
i\tau {\bf k}^2 -2i\tau e {\bf k}\cdot \bepsilon \int_{\xi'}^\xi A \right)      
$$                                                                              
\be\label{trw}  
= {-i \over 4\pi \tau} e^{-i {({\bf y}-{\bf y}')^2 \over 4\tau} -i\tau e^2      
\left({1 \over \xi-\xi'} \int_{\xi'}^\xi A                                      
\right)^2 } {\rm exp}\left(ie{({\bf y}-{\bf y}')                                
\cdot \bepsilon \over \xi-\xi'} \int_{\xi'}^\xi A \right)                       
.\ee 
This agrees with Schwinger's result, the last term just                         
giving the rest of the four-dimensional 
Wilson line, \ie using                                   
$$
\int_{x'_\mu}^{x_\mu} dz^\nu =
 {(x^\nu - x'^\nu) \over \xi-\xi'} \int_{\xi'}^{  
\xi} d{\bar \xi}                                                                
$$
for a
straight-line
integral,   
we get     
$$
     -e
\int_{x' \rightarrow x} A = -e \varepsilon^0 \int_{\xi'}^\xi d{\bar \xi}   
   A({\bar
\xi}) + e{({\bf y}-{\bf y}')     \cdot \bepsilon
\over \xi-\xi'} \int_{\xi'}^\xi d{\bar\xi} A({\bar \xi})        
,$$ 
giving  the two pieces (\ref{lw}) and (\ref{trw}) of the full Wilson line.

\subsubsection*{Effective
Mass}

For a
periodic
wave
$A(\xi) =
A\,{\rm
sin}\,(\omega\xi)$
one can define an effective mass describing the effects of the plane wave
on the inertial properties of the particle. From the general expression in
(\ref{ddf}) we
read off the effective mass for propagation in the $(t,x)$ plane, as a function
of the transverse momentum ${\bf k}$, given by
\be
m^2_{\rm eff}({\bf k})
 = m^2 +e^2 \lim_{|\xi-\xi'|\to \infty} \int_{\xi'}^{\xi} {
\CA_{\bf k}^2 \over \xi-\xi'}= m^2 + e^2 {\overline{\CA_{\bf k}^2}}
.\ee
In the commutative case ${\overline{\CA_{\bf k}^2}}=
{\overline {A^2}} = A^2/2$ and the effective mass is
actually independent of ${\bf k}$. For the non-commutative case, we find exactly
the same result for charge-one particles:
\be
m^2_{\rm eff}({\bf k})_{q=1} = m^2 +\shalf e^2 A^2
,\ee
whereas the neutral particles show $\theta$-dependent resonant effects:
\be
 m^2_{\rm eff}({\bf k})_{q=0} = m^2 +2 e^2 A^2\,{\rm sin}^2
 \left(\shalf\omega n\theta k\right).
\ee                                                                            
Namely, the effective mass renormalization can be completely cancelled if
a tuning is made of the transverse momentum and the non-commutative deformation
parameter.

\subsubsection*{Effective
Action}
       The one-loop  effective action
   $\delta \CL$ is defined by:                                        
\be \label{efwave}                                                                             {\rm
exp}\left(i\int \delta
 \CL \right) = {{\rm det} (i\Dirac_{A\star} -m) \over {\rm det} (
i\dirac-m)}   ={{\rm
det}^{\half}
(H(\Dirac_{A\star}))
\over {\rm
det}^{\half}(-\pt^2
-m^2)}                                                                    
.\ee    
This quantity was computed by Schwinger in the commutative case and
shown to be trivial, in agreement with the expectations from the
point of view of effective field theory. Namely all local Lorentz-invariants 
constructed from the field strength $F_{\mu\nu}$ or its dual $F^*_{\mu\nu}$ 
 vanish in the plane-wave background. In fact, it is not hard to show that
all polynomial  invariants constructed from powers  $\Fhat, \Fhat^*$ and
non-commutative covariant derivatives also vanish in the non-commutative
plane-wave background. Thus, we expect the effective action     
 (\ref{efwave}) to vanish, up to boundary terms.

The  proper-time representation is: 
\be                             
   \delta
\CL(x) = -{i\over 2} \int_{-\infty}^{0} {d\tau \over \tau} \,{\rm tr}_{
   \rm Dirac}
\;U(x;x,\tau) - (\bepsilon =0)                                       
,\ee    
                        where the
$m^2-i0$ prescription is assumed. We can get rid of it       
   by rotating the
contour to euclidean proper-time  $\tau =is $,              
     so
that                                                                         
\be                                                                              \delta
\CL(x) = {i\over 2} \int_0^\infty {ds \over s}\,{\rm tr}_{                   \rm Dirac}
\;U(x;x,is) - (\bepsilon =0)                                         
.\ee
Evaluating the kernel (\ref{bosk})
 in the limit $x\rightarrow x'$, $\xi\rightarrow \xi'$      we
get                                                                          
\be                                                                               U(x;x,is)
= {i\over 4\pi s} e^{-s\left(m^2  + {e\over 2} f_{\mu\nu}             
 \sigma^{\mu\nu}
 \,\CA_{\bf k}'(\xi)\right)} \int {d{\bf k} \over (2\pi)^2}             
 {\rm exp}\left[-s\left({\bf k} - e                                             
\bepsilon \CA_{\bf k} (\xi)\right)^2
 \right]                                             
.\ee 
Using  that, for a plane wave                                                   
\be 
{\rm tr}_{\rm Dirac} e^{-s {e\over 2} f\cdot \sigma \CA'} = 4                   
,\ee
we find the final
result for the effective action:  
\be \label{effaction}  
\int \delta
\CL(x) = -{1\over 2\pi} \int_0^\infty
 {ds\over s^2} e^{-sm^2} \int d^4 x \,
\int        {d{\bf k} \over
(2\pi)^2} {\rm exp}\left[-s\left({\bf k}  - e                    \bepsilon
\CA_{\bf k} (\xi)\right)^2 \right]    -(\bepsilon=0).   
\ee         

We can analyze the effective action (\ref{effaction}) more carefully to
show that indeed it will only lead to boundary terms. Surprisingly, the
result depends once again on the celebrated IR/UV correspondence of
\cite{uvir}. To define the effective action derived from (\ref{effaction})
we need to specify a way to regularize the $x$- and ${\bf k}$-integrals. We
will assume the wave-front to be compact supported with a size $\Delta$. Using
light-cone variables $\xi=t-x, \eta=t+x$, the volume measure becomes 
$d\xi\,d\eta\,d{\bf y}$. We will consider the space-time integral to be
done in a space-time box of volume $L^4$ (assuming for simplicity $L\gg 
\Delta$). Since $\CA_{\bf k}(\xi)$ is a pulse centered at $\pm \vert n\theta k
\vert$
 (for $q=1$) or two pulses centered  one at $\vert n\theta k \vert$
 and the other at  
$-\vert n\theta k \vert$, the momentum integration variable
 is constrained to satisfy $\vert n\theta k\vert 
\leq L$.  Once again the presence of $\theta$ translates an infrared
cutoff $L$ in $x$ into an ultraviolet cutoff in ${\bf k}$. 

For each value of ${\bf k}$, the integral over $\xi$ can be done, since
the argument depends only on $\xi$. If we
are away from the boundary of the box, the value of the integral near
$\xi =-{L\over 2}$ is
$$
C_1 + \xi\,e^{-s{\bf k}^2},
$$
while near $\xi={L\over 2}$ it is
$$
C_2 + \xi\,e^{-s{\bf k}^2}
$$
and the integral is just the difference between both expressions evaluated
at $\pm {L\over 2}$:
\be
\label{finac}
(C_2 - C_1) + L\,e^{-s{\bf k}^2}.
\ee
The value of $C_2 -C_1$ depends on the particular shape of the pulse and on 
$s$, but in the limit we are considering, where $\Delta \ll L$, they are
slowly varying functions of ${\bf k}$. When we perform the integral over 
${\bf k}$, the first term produces a quadratic divergence $\sim (C_2 - C_1)\,
(L/\theta)^2$ as $L\rightarrow \infty$, where $\theta$ is the typical
eigenvalue of the $\theta$-matrix. This exhibits a mild version of the
IR/UV mixing that appears in  loop computations in \cite{uvir, uvirs}.

We see that a careful correlation of ultraviolet and infrared cutoffs, such
that the shifted pulses are always `inside the box', guarantees 
that (\ref{effaction}), being a function of a single space-time variable, is
a total derivative.   Hence
we conclude as in the commutative case that there is no bulk contribution
to the effective action.


\subsection*{Acknowledgements}

We are indebted to Costas Bachas,  
Esperanza L\'opez, Quim Gomis, Antonio Gonz\'alez-Arroyo,
Karl Landsteiner, Eliezer Rabinovici and Miguel A. V\'azquez-Mozo for
useful discussions. L. A-G. would like to   thank the Physics Department
at Berlin's Von Humboldt University for hospitality, where part of this
work was done. J.L.F.B. would like to thank the physics department of
the University at Paris-VII, where part of this work was done.

\end{document}